\documentclass[galaxies,communication,submit,pdftex]{Definitions/mdpi} 

\firstpage{1} 
\pubvolume{11}
\issuenum{3}
\articlenumber{0}
\pubyear{2023}
\copyrightyear{2023}
\datereceived{17 April 2023} 
\dateaccepted{5 July 2023} 
\datepublished{} 
\hreflink{https://doi.org/} 
\pdfoutput=1

\usepackage{threeparttable}

\Title{Probing neutrino production in blazars by millimeter VLBI}

\TitleCitation{Probing blazars and neutrino}

\Author{Y. Y. Kovalev$^{1,2}$\orcidA{}, A. V. Plavin$^3$\orcidB{}, A. B. Pushkarev$^{4,3}$\orcidD{}, S. V. Troitsky$^{5,6}$\orcidC{}}

\AuthorNames{Y. Y. Kovalev, A. V. Plavin, A. B. Pushkarev, S. V. Troitsky}

\AuthorCitation{Kovalev et al.}

\address{%
$^1$ \quad Max-Planck-Institut f\"ur Radioastronomie, Auf dem H\"ugel 69, 53121 Bonn, Germany\\
$^2$ \quad Black Hole Initiative at Harvard University, 20 Garden Street, Cambridge, MA 02138, USA\\
$^3$ \quad Lebedev Physical Institute of the Russian Academy of Sciences, Leninsky prospekt 53, 119991 Moscow, Russia\\
$^4$ \quad Crimean Astrophysical Observatory, 298409 Nauchny, Crimea, Russia\\
$^5$ \quad Institute for Nuclear Research of the Russian Academy of Sciences, 60th October Anniversary Prospect 7a, Moscow, Russia\\
$^6$ \quad Faculty of Physics, M.V. Lomonosov Moscow State University, 1-2 Leninskie Gory,  Moscow 119991, Russia
}

\corres{Correspondence: yykovalev@gmail.com, alexander@plav.in}

\abstract{
The advancement of neutrino observatories has sparked a surge in multi-messenger astronomy. Multiple neutrino associations among blazars are reported while neutrino production site is located within their central (sub)parsecs. Yet many questions remain on the nature of those processes. The next generation Event Horizon Telescope (ngEHT) is uniquely positioned for these studies, as its high frequency and resolution can probe both the accretion disk region and the parsec-scale jet. This opens up new opportunities for connecting the two regions and unraveling the proton acceleration and neutrino production in blazars. We outline observational strategies for ngEHT and highlight what it can contribute to the multi-messenger study of blazars.
}


\begin{document}

\section{Introduction: current status of high-energy neutrino studies, blazars --- neutrino connection}
\label{s:intro}

Neutrino observatories, such as IceCube, ANTARES (Astronomy with a Neutrino Telescope and Abyss environmental RESearch project), and Baikal-GVD (Gigaton Volume Detector) have been convincingly detecting astrophysical neutrinos at TeV to PeV energies \cite{IceCubeFirst26,2016ApJ...833....3A,2018ApJ...853L...7A,BaikalDiffuse}. Despite these observations, little was known about the origin of energetic astrophysical neutrinos until recently.

\begin{figure*}
    \centering
    \includegraphics[width=0.9\linewidth]{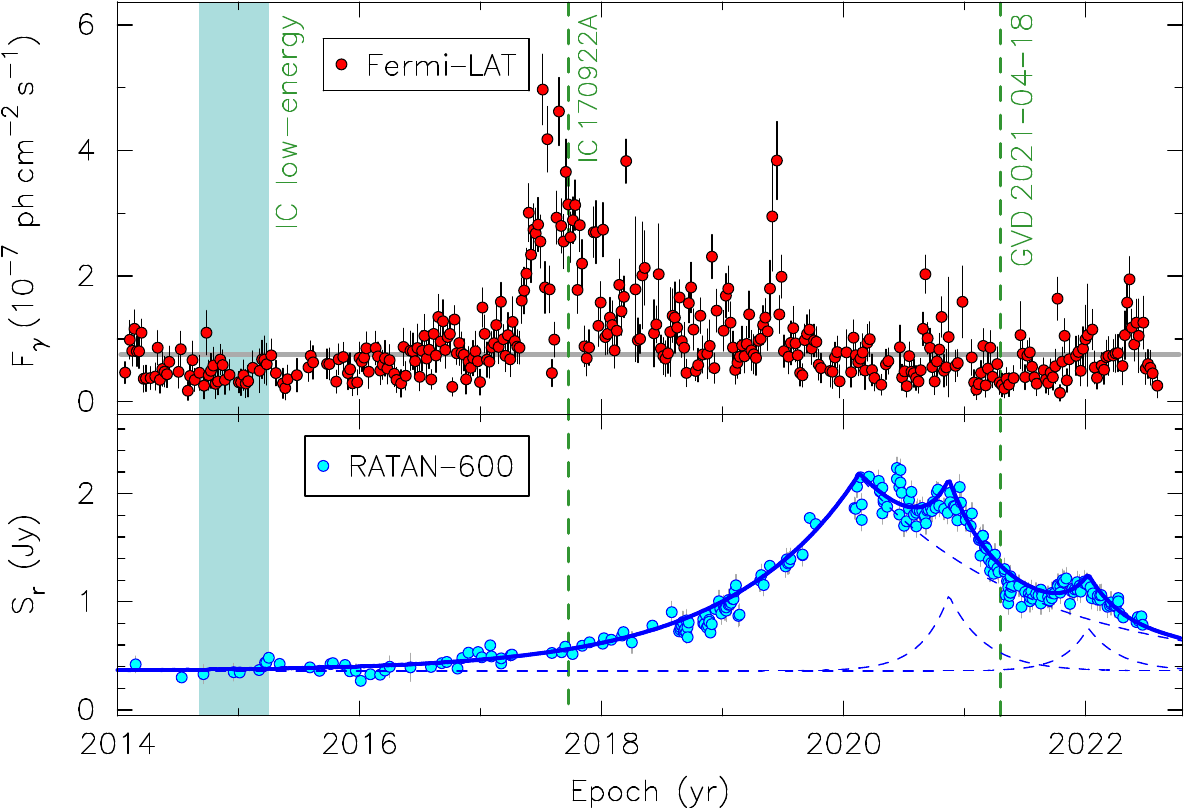}
    \caption{
    Radio and gamma-ray light curves of TXS~0506+056. Top: \textit{Fermi} LAT weekly binning light curve \citep{2023ApJS..265...31A} of the gamma-ray source 4FGLJ0509.4+0542 positionally associated with the quasar TXS~0506+056. The gray horizontal line denotes the median gamma-ray flux. Bottom: RATAN-600 light curve at 11~GHz. The radio light curve is decomposed by three radio flares depicted by the blue dashed lines, the sum of which is represented by a thick line \citep{Baikal0506}. The vertical lines denote the times of neutrino arrival \cite{IceCubeTXSgamma,Baikal0506}.}
    \label{f:0506_lc}
\end{figure*}

Blazars, a class of active galactic nuclei (AGN), have been considered as potential neutrino sources since the very early days of multi-messenger astronomy \cite{BerezinskyNeutrino77}. Observational evidence for a blazar-neutrino connection started to emerge in recent years. First, the blazar TXS~0506+056 was associated with a high-energy neutrino, which coincided with a gamma-ray flare in 2017 \cite{IceCubeTXSgamma}. This association was in contrast with a lack of systematic connection between gamma-ray-loud blazars and neutrinos (see, e.g.~\cite{corr-1611.06338Neronov,Murase10percent}). Then, numerous radio-bright blazars were shown to emit neutrinos with energies from TeVs to PeVs \cite{Kadler16,neutradio1,neutradio2,hovatta2021,Illuminati2021,Baikal2021,neutradio3,2023arXiv230511263B}. The detection of this correlation is driven by the unique capabilities of very-long-baseline interferometry (VLBI): this is the only technique able to directly probe and resolve central (sub)parsecs in AGNs at cosmological distances. Blazars emit neutrinos preferentially at the times of their flares (\autoref{f:0506_lc}) visible in radio bands \cite{neutradio1,hovatta2021,2021MNRAS.503.3145B,neutradio3}. Still, the neutrino production mechanism and the physical regions where it occurs remain unclear. The observed connection of neutrinos with radio emission from compact jet regions emphasizes the importance of high-resolution studies in answering these questions. VLBI is the best direct visual evidence we can obtain in astronomy.

For a general discussion of multi-wavelength and multi-messenger studies with the ngEHT, see \citet{2023Galax..11...17L}.
In this paper, we present the progress in the multi-messenger astronomy studies of cosmic neutrinos, their probable association with blazars, challenges and a critical role to be played by ngEHT \citep{2021ApJS..253....5R,ngEHT_KSG,2023arXiv230608787D} in addressing exciting open questions of high-energy neutrino production.

\section{Neutrino production in blazars: open questions
}

Assuming no particle physics beyond the Standard Model, astrophysical neutrinos with energies above TeV can be produced only in interactions of relativistic hadrons~--- protons or nuclei~--- with ambient matter or radiation, see, e.g.~\cite{ST-UFN} for a recent review. This fits well the observational evidence discussed in \autoref{s:intro} because the nonthermal radiation of blazars gives a clear signal that particles are accelerated there. However, both the amount of relativistic hadrons in AGN, and the degree to which these hadrons contribute to the observed electromagnetic radiation, are uncertain. Population studies suggest \cite{NeronovEvolution,Finley-2005.02395,ST-UFN} that their contribution is small, and neutrino luminosities of blazars are orders of magnitude lower than photon luminosities. Consequently, one may imagine neutrino production in various places in a blazar and by means of different mechanisms. 

The main challenge is to explain the production of neutrinos of very different energies, from a few TeV \cite{IceCube:TXSlow,neutradio2} to sub-PeV \cite{IceCubeTXSgamma, neutradio1}, in sources of the same class. For the $p\gamma$ mechanism, expected to dominate in blazars \cite{Boettcher-rev}, the wide neutrino energy range requires the presence of target photons with a very broad distribution of energies. Conventional models of high-energy neutrino production in AGN, known for decades, e.g.\ \cite{BerezGinzb,Eichler1979,Stecker:1991vm}, as well as their modern versions, e.g.\ \cite{NeronovWhich,Stecker-1305.7404,Kalashev:2014vya}, often experience problems in explaining the lower-energy part of the observed neutrino flux, in particular because the target photons from the accretion disk are expected to have energies $\sim (10 \dots 100)$~eV, while $\sim 10$~keV are required for intense production of $\sim 10$-TeV neutrinos. 

While neutrinos have already been associated with VLBI-bright blazars \cite{neutradio2,neutradio3} and with their radio flares \cite{neutradio1,hovatta2021}, these results were based on observations at centimeter wavelengths. There, synchrotron self-absorption prevents one from detailed spatio-temporal studies of the AGN central sub-parsec parts \citep[e.g.,][]{2017A&ARv..25....4B}. 
To summarize, the open questions of the blazar-neutrino astrophysics are the following: (i) how are protons accelerated, (ii) what is the neutrino production process, $p\gamma$ or $pp$, (iii) from where do seed (X-ray) photons originate in case of $p\gamma$, (iv) where are neutrinos produced?
Note that (ii) and (iv) can be different, and multizone models may be required to explain all observations consistently.

\section{Neutrino astronomy in the ngEHT era}

Currently, studies of high-energy astrophysical neutrinos and their sources are limited by the sensitivity and resolution of neutrino observatories. The situation is rapidly changing, as their capabilities are increasingly improving. The next-generation IceCube-Gen2 will grow the telescope volume tenfold, from 1 to 10 cubic kilometers, aiming at a corresponding increase in detection rates by 2033 \cite{IC_Gen2}. The Baikal-GVD detector has already reached the effective volume of 0.5 cubic kilometer and continues to grow and improve event reconstruction algorithms \cite{2022icrc.confE...2B}. KM3Net (Cubic Kilometre Neutrino Telescope), a neutrino observatory in the Mediterranean, is being constructed and has already started yielding its first results \cite{2019APh...111..100A}. Together, these instruments will provide a qualitative leap in both the number of detected astrophysical neutrinos and in the precision of their localization.

An increasing number of well-localized neutrinos will lead to a reliable identification of individual blazars as neutrino sources. Moreover, it should be possible to highlight specific time periods with more prominent neutrino emission. This brings new challenges and opportunities to the EM counterpart of such multi-messenger studies. 

The planned ngEHT array \citep{2021ApJS..253....5R,ngEHT_KSG,2023arXiv230608787D} will provide superior angular resolution, dynamic range and sensitivity in Stokes I and polarization at 3, 1.3, and 0.7~mm. 
This will allow scientists to observe and monitor the central (sub)parsecs of neutrino-emitting blazars at the highest resolution and frequency possible, significantly alleviating the synchrotron opacity problem of the current centimeter-wavelength VLBI.
The ngEHT will be able to probe both the accretion disk region and the parsec-scale jet base, opening new opportunities for connecting the two regions and unraveling the proton acceleration and neutrino production in blazars.

\section{Planning ngEHT experiment}

\begin{table}
\caption{Most probable neutrino candidates among VLBI-selected bright blazars.}
\label{t:assoc}
\centering
\begin{tabular}{llcclc}
\hline\hline
\multicolumn{2}{c}{Blazar name} & $z$ & $S_\textrm{86 GHz}^\textrm{VLBI}$  & \# of high-energy  & Reference\\
B1950      & Alias     &      & (Jy)           & neutrinos (and dates)& \\
(1)        & (2)       & (3)  & (4)            & (5)       & (6)\\
\hline
0506$+$056 &           & 0.34 &  0.6$^\dagger$ & 2 (2017-09-22, 2021-04-18)        & \citep{IceCubeTXSgamma,Baikal0506}\\
0735$+$178 & OI~158    & 0.45 &  0.6           & 1-4 (2021-12-04\&08)      & \citep{neutradio3,2008AJ....136..159L}\\
1253$-$055 & 3C\,279   & 0.54 & 22.7           & 1 (2015-09-26)        & \citep{neutradio1,2019AA...622A..92N}\\
1502$+$106 & OR~103    & 1.84 &  0.6           & 1 (2019-07-30)        & \citep{neutradio1,2008AJ....136..159L}\\
1730$-$130 & NRAO\,530 & 0.90 &  1.9$^\dagger$ & 1 (2016-01-28)        & \citep{neutradio1,2010ApJS..189....1A}\\
1741$-$038 &           & 1.05 &  3.2           & 2 (2011-09-30, 2022-02-05)        & \citep{neutradio3,2008AJ....136..159L}\\
1749$+$096 & OT~081    & 0.32 &  2.4           & 1 (2022-03-03)        & \citep{neutradio3,2008AJ....136..159L}\\
2145$+$067 &4C\,$+$06.69& 1.00 &  3.6$^\dagger$ & 1 (2015-08-12)        & \citep{neutradio1,2010ApJS..189....1A}\\
\hline
\end{tabular}
\begin{tablenotes}
\item
Notes:
Publications that selected each blazar as highly probable neutrino emitter and measured their flux densities are referenced in the last column. The 0506+056 blazar was the first and only to be distinguished by the IceCube, while the others in this list were found by statistical analysis of complete VLBI-selected samples. The dates for high-energy neutrinos are shown in the format YYYY-MM-DD.
\item
$^\dagger$Estimated from nearby VLBI measurements at 15 and 43~GHz of MOJAVE and Boston university VLBA programs.
\end{tablenotes}
\end{table}

Below, we discuss several approaches to study and understand the physics behind the connection between neutrino production and EM activity from the jet upstream to the vicinity of the central engine --- a possibility which will be realized by ngEHT.
Before elaborating on observing campaigns, we note the following important complications of neutrino astrophysics that affect suggested scenarios below.
A typical probability of a neutrino with an energy above 100~TeV to be of an astrophysical origin is around 50\%, and it drops significantly for lower energies \citep{AhlersHalzen,IceCat-1}. 
A typical 90\% error region of a highly probable high-energy neutrino is several square degrees \citep{IceCubeOldAlerts,IceCat-1}. 
Some neutrinos might arrive from nearby non-jetted AGNs \citep{2022Sci...378..538I} or even from our Galaxy and its relativistic objects \citep{2022ApJ...940L..41K,2021PhyU...64.1261T,Bykov:2021maf,2023MNRAS.tmpL..80K,2023ApJ...949L..12A,IC_Galaxy2023}.
On top of that, we know very little about mechanisms of neutrino production in blazars --- so there is no streetlight under which we can plan our search.

We expect that a variety of blazars would be associated with neutrinos and allow us to select optimal ngEHT targets taking both physical properties and technical or observational limitations into account.
Within the current understanding and the experience accumulated from observational searches for high-energy neutrino counterparts, the following three scenarios for monitoring observations are suggested.

\textbf{Scenario~1:} Observations of blazars associated with selected new high-energy neutrino alerts immediately after neutrino arrival. Up to several blazar-associated high-energy alerts per year are expected. When two or three neutrino telescopes become fully operational, one might conservatively require two alerts for a given target to arrive within several days.\\
Pros: most efficient since linked to a specific event.
Cons: will only be able to probe the state of an associated object after neutrino arrival.

\textbf{Scenario~2}: Observations of a sample of selected blazars reliably identified previously as neutrino sources. See Table~\ref{t:assoc} with most probable neutrino candidates to date.\\
Pros: optimal in terms of observed sample and complete temporal coverage of events.
Cons: so far, a very limited number of cases is known with repeated neutrino detection from the same source (Table~\ref{t:assoc}, column 5), but their list can grow.

\textbf{Scenario~3:} Observations of a complete VLBI-flux-density limited sample of 50-100 brightest blazars with 3~mm VLBI flux density above 1~Jy \citep{2008AJ....136..159L,2019AA...622A..92N}.\\
Pros: full temporal coverage of expected events, possibility to compare neutrino-emitting and neutrino-non-emitting blazars calculating robust significance of a coincidence \citep{neutradio3,2022A&A...666A..36L}; an option to combine such observations with other ngEHT science cases \citep{ngEHT_KSG}.
Cons: observationally expensive.

Tracing changes in the compact structure of blazars during and around periods of increased neutrino emission requires multi-epoch monitoring at a high resolution provided by ngEHT. 
To roughly estimate required observing time, we expect that one imaging epoch per target takes 4-8~hours.
The observations should happen with a cadence between two weeks and one month (an estimate based on experience gained by the 7~mm blazar VLBA monitoring program \citep{2022ApJS..260...12W}) and produce polarization images with Stokes I dynamic range about or better than 1000:1, preferably multi-frequency with a possibility for Faraday rotation measure (RM) and spectral analysis. 
From this, we will be able to constrain the following source properties.
\begin{enumerate}
    \item Jet kinematics measurements will allow us to better estimate Doppler boosting and jet viewing angle following \citep[e.g.,][]{2021ApJ...923...67H,2022ApJS..260...12W}, constrain plasma acceleration \citep[e.g.,][]{2015ApJ...798..134H,2022ApJS..260...12W}.
    Jet geometry profile studies will constrain jet formation and collimation \citep{2012ApJ...745L..28A,2020MNRAS.495.3576K}.
    \item Jet kinematics will also deliver information about newborn jet features \citep[e.g.,][]{2021ApJ...923...30L,2022ApJS..260...12W},
    measure ejection epochs of features possibly associated with neutrino events, compare those with neutrino arrival time and locate the neutrino production zone from the measured delay.
    Compare with similar analysis for VLBI-$\gamma$-ray studies \citep{2017ApJ...846...98J,2022MNRAS.510..469K}.
    \item Faraday RM, reconstructed EVPAs and analysis of radio spectrum together with core-shift measurements will deliver information on the magnetic field structure, its strength and its changes \citep[e.g.,][]{1998A&AS..132..261L,2017MNRAS.467...83K,2015Sci...348..311M}, which might be related to physical conditions required for neutrino production.
    \item Monitoring overall changes in the millimeter parsec- and sub-parsec scale structure of blazars at extreme resolution of ngEHT will allow us to distinguish between flares in disks and in jets \citep[e.g.][]{Murase-rev,2022arXiv221203151K} related to neutrino production if resolution, sensitivity, and opacity permit. Observing in this regime, we will be able to overcome significant delays related to synchrotron self-absorption at lower radio frequencies (see \autoref{f:0506_lc} as well as \citep{2022MNRAS.510..469K}).
\end{enumerate}

We underline that studying a complete sample of AGN with understandable properties will allow us not only to relate the observed changes to detected neutrinos but also set a robust significance on that association, following the approach suggested by  \citet{neutradio3}.

\section{Synergy with other facilities}

The Square Kilometer Array \citep[SKA,][]{SKA_VLBI} and especially the next generation Very Large Array \citep[ngVLA,][]{2018ASPC..517....3M,2018ASPC..517...15S} going as high as 100~GHz will allow for monitoring much larger samples of VLBI-selected AGN as well as fast imaging of fields of neutrino arrival, pre-selecting most probable neutrino candidates for ngEHT targeted studies.
Wide-field telescopes like 
the optical Legacy Survey of Space and Time \citep[LSST,][]{2019ApJ...873..111I} will allow scientists to much better associate blazars with neutrinos in cases if flaring activity is confirmed as a valid indicator \citep[e.g.,][]{2018Sci...361.1378I,2020ApJ...896L..19L,neutradio3}.
Moreover, optical and UV telescopes can separate flares happening in jets and accretion disks, analyzing optical color and polarization.
Seed photons are expected from X-rays \citep{neutradio2,Murase-rev,neutXray}, which is where current and new generation space X-ray telescopes will be very helpful. High energies \citep[e.g., the Cherenkov Telescope Array -- CTA,][]{2011ExA....32..193A,2019scta.book.....C} will continue supporting the gamma-ray/TeV -- neutrino analysis and allow checking if neutrino production zone is actually opaque to gamma-rays.

\section{Summary}

The ngEHT will revolutionize VLBI imaging capabilities by bringing together the power of surpassing resolution, advanced dynamic range, and sensitive polarization data. What makes it unique, however, is its remarkable immunity to synchrotron absorption.
It will allow probing regions all the way from the accretion disk to the parsec-scale jet \citep[e.g.,][]{2023Natur.616..686L} and study the most probable sources of high-energy neutrinos --- blazars.
By the time of ngEHT operations, three large high-energy neutrino telescopes will be fully functional: IceCube, KM3NeT, and Baikal-GVD.
This paper formulates the science case, presents eight most probable associations to date, and suggests observational strategies to address very exciting and wide open questions of proton acceleration and neutrino production.

\funding{
This publication is funded in part by the Gordon and Betty Moore Foundation. It was also made possible through the support of a grant from the John Templeton Foundation. The opinions expressed in this publication are those of the author(s) and do not necessarily reflect the views of these Foundations.
}

\acknowledgments{
We thank the ngEHT team for discussions, Eduardo Ros as well as anonymous referees for useful comments on the manuscript, and Elena Bazanova for language editing.
%
The VLBA is an instrument of the National Radio Astronomy Observatory. The National Radio Astronomy Observatory is a facility of the National Science Foundation operated by Associated Universities, Inc.
}

\conflictsofinterest{The authors declare no conflict of interest.} 

%

\clearpage

\newcommand\aap{A\&A}                
\let\astap=\aap                          
\newcommand\aapr{A\&ARv}             
\newcommand\aaps{A\&AS}              
\newcommand\actaa{Acta Astron.}      
\newcommand\afz{Afz}                 
\newcommand\aj{AJ}                   
\newcommand\ao{Appl. Opt.}           
\let\applopt=\ao                         
\newcommand\aplett{Astrophys.~Lett.} 
\newcommand\apj{ApJ}                 
\newcommand\apjl{ApJ}                
\let\apjlett=\apjl                       
\newcommand\apjs{ApJS}               
\let\apjsupp=\apjs                       
\newcommand\apss{Ap\&SS}             
\newcommand\araa{ARA\&A}             
\newcommand\arep{Astron. Rep.}       
\newcommand\aspc{ASP Conf. Ser.}     
\newcommand\azh{Azh}                 
\newcommand\baas{BAAS}               
\newcommand\bac{Bull. Astron. Inst. Czechoslovakia} 
\newcommand\bain{Bull. Astron. Inst. Netherlands} 
\newcommand\caa{Chinese Astron. Astrophys.} 
\newcommand\cjaa{Chinese J.~Astron. Astrophys.} 
\newcommand\fcp{Fundamentals Cosmic Phys.}  
\newcommand\gca{Geochimica Cosmochimica Acta}   
\newcommand\grl{Geophys. Res. Lett.} 
\newcommand\iaucirc{IAU~Circ.}       
\newcommand\icarus{Icarus}           
\newcommand\japa{J.~Astrophys. Astron.} 
\newcommand\jcap{J.~Cosmology Astropart. Phys.} 
\newcommand\jcp{J.~Chem.~Phys.}      
\newcommand\jgr{J.~Geophys.~Res.}    
\newcommand\jqsrt{J.~Quant. Spectrosc. Radiative Transfer} 
\newcommand\jrasc{J.~R.~Astron. Soc. Canada} 
\newcommand\memras{Mem.~RAS}         
\newcommand\memsai{Mem. Soc. Astron. Italiana} 
\newcommand\mnassa{MNASSA}           
\newcommand\mnras{MNRAS}             
\newcommand\na{New~Astron.}          
\newcommand\nar{New~Astron.~Rev.}    
\newcommand\nat{Nature}              
\newcommand\nphysa{Nuclear Phys.~A}  
\newcommand\pra{Phys. Rev.~A}        
\newcommand\prb{Phys. Rev.~B}        
\newcommand\prc{Phys. Rev.~C}        
\newcommand\prd{Phys. Rev.~D}        
\newcommand\pre{Phys. Rev.~E}        
\newcommand\prl{Phys. Rev.~Lett.}    
\newcommand\pasa{Publ. Astron. Soc. Australia}  
\newcommand\pasp{PASP}               
\newcommand\pasj{PASJ}               
\newcommand\physrep{Phys.~Rep.}      
\newcommand\physscr{Phys.~Scr.}      
\newcommand\planss{Planet. Space~Sci.} 
\newcommand\procspie{Proc.~SPIE}     
\newcommand\rmxaa{Rev. Mex. Astron. Astrofis.} 
\newcommand\qjras{QJRAS}             
\newcommand\sci{Science}             
\newcommand\skytel{Sky \& Telesc.}   
\newcommand\solphys{Sol.~Phys.}      
\newcommand\sovast{Soviet~Ast.}      
\newcommand\ssr{Space Sci. Rev.}     
\newcommand\zap{Z.~Astrophys.}       

\begin{adjustwidth}{-\extralength}{0cm}
\reftitle{References}
\bibliography{neutradio}
\end{adjustwidth}

\end{document}